\documentclass[prl,twocolumn,showpacs,preprintnumbers,amsmath,amssymb]{revtex4-1}
\usepackage{dcolumn}
\usepackage{color}

\usepackage{graphicx}
\usepackage{amssymb}
\usepackage{amsmath}
\usepackage{bm}
\usepackage{epsfig}
\usepackage{ulem}

\newcommand {\be}{\begin{equation}}
\newcommand {\ee}{\end{equation}}
\newcommand {\bea}{\begin{eqnarray}}
\newcommand {\eea}{\end{eqnarray}}
\newcommand {\FIG}[1]{Fig. \ref{#1}}

\newcommand {\EQ}[1]{Eq. (\ref{#1})}

\newcommand {\PRE}[1]{{Phys. Rev. E} {\bf {#1}}}

\newcommand {\PRL}[1]{{Phys. Rev. Lett.} {\bf {#1}}}

\newcommand {\EPJB}[1]{{Eur. Phys. J. B} {\bf {#1}}}
\newcommand {\NAT}[1]{{Nature} {\bf {#1}}}

\newcommand {\SCI}[1]{{Science} {\bf {#1}}}

\begin{document}

\title{Explosive site percolation with a product rule}
\author{Woosik Choi}
\author{Soon-Hyung Yook}\email{syook@khu.ac.kr}
\author{Yup Kim} 
\affiliation{Department of Physics and Research Institute for Basic
Sciences, Kyung Hee University, Seoul 130-701, Korea}
\date{\today}
\begin{abstract}
We study the site percolation under Achlioptas process (AP) with a
product rule in a $2-dimensional$ (2D) square lattice. From the
measurement of the cluster size distribution, $P_s$, we find that
$P_s$ has a very robust power-law regime followed by a stable hump
near the transition threshold. Based on the careful analysis
on the $P_s$ distribution, we show that the transition should be
discontinuous. The existence of the hysteresis loop in order
parameter also verifies that the transition is discontinuous in
2D. Moreover we also show that the transition nature from the
product rule is not the same as that from a sum rule in 2D.
\end{abstract}


\maketitle

The percolation transition describing the emergence of large-scale
connectivity in lattice systems or complex networks has been
extensively studied in statistical mechanics and related fields
due to its possible applications to various phenomena such as
sol-gel transition and polymerization, resistor networks, and
epidemic spreading \cite{Stauffer_book}. When the occupation
probability of node (site) is lower than certain threshold $p_c$,
all the clusters are microscopic. As the occupation probability
increases, the macroscopically connected cluster emerges. Such
transition in the ordinary percolation is a continuous transition
\cite{Stauffer_book}.

On the other hand, there have been several attempts to find a
percolation model which undergoes a discontinuous transition. The
discontinuous percolation transition can be found in the modelling
of magnetic systems with significant competition between exchange
and crystal-field interactions \cite{Andersen76,Chalupa79}. The
similar phenomena has been found in financial systems
\cite{Kim08}, in which two equally probable phases exit. Other
examples of the discontinuous transition in percolation are the
formation of infinite cluster under a central-force
\cite{Moukarzel97} and the cascade of failure in interdependent
networks \cite{Buldyrev10}.

Recently, Achlioptas {\it et al}. \cite{Achlioptas09} suggested a
simple process in which the growth of large clusters is
systematically suppressed and the process is usually called as
Achlioptas process (AP). Based on the analysis of transition
interval it was argued that the percolation transition under AP is
{\it explosive} and discontinuous. Several variant of models have
been investigated to understand the general properties and
conditions which cause such non-trivial discontinuous transition
\cite{Ziff09,Cho09,Radicchi,YKim10}. Some examples of such
non-trivial transition has been found in nano-tube based system
\cite{YKim10}, protein homology network \cite{Rosenfeld10}, and
community formation \cite{Pan10}.

However, more recent studies on the percolation transition under
AP reveals several evidences which strongly suggest that the
transition can be continuous. For example, da Costa {\it et al}.
\cite{daCosta10} argued that the transition in the complete graph
(CP) is continuous, even though the order parameter exponent 
is very small ($\beta\simeq 0.056$). From the measurement of
the cluster size distribution Lee {\it et al}. \cite{Lee11} also argued
that the transition in CP is continuous. Grassberger {\it et al}.
\cite{Grassberger11} also argued that the transition, even in the
low-dimensional systems, can be continuous based on a measurement
of the order parameter distribution.

Since most of the studies on the criticality of AP process are
restricted to the infinite dimensional systems, it is still not
clear whether AP also produces a continuous transition in lower
dimensional systems or not. For example, in the bond percolation
under AP in a 2-dimensional (2D) square lattice, the product rule
was argued to produces a discontinuous transition based on a
finite-size scaling \cite{Ziff09,Araujo11}. In contrast Grassberger
{\it et al}. \cite{Grassberger11} was argued that transition of the
2D AP bond percolation is still continuous. We therefore
cannot exclude the possibility that the transition nature of the
AP in the mean-field limit can be different from that in
lower-dimensional systems, like the Potts model \cite{Wu82}.
Moreover, based on the measurement of hysteresis \cite{Bastas11}, a
sum rule for the 2D site percolation possibly makes the transition
continuous in the thermodynamic limit. This indicates that under
the AP-like processes the bond percolation and site percolation
may have different transition natures in the 2D lattice. In the
ordinary percolation, bond and site percolations  are known to
belong to the same universality class \cite{Stauffer_book}. In
contrast, the results in Refs. \cite{Ziff09,Araujo11,Bastas11}
show the possibility that under AP the bond percolation with a
product rule and the site percolation with a sum rule do not
belong to the same universality class. Therefore, it is
theoretically important and interesting to investigate
whether in a low-dimensional system the product rule and the sum
rule belong to the same universality class or not. In order to
achieve this purpose, we investigate the site percolation under AP
with a product rule and show that AP with the product rule
produces a clear discontinuous transition in a 2D lattice. For
this we carefully analyze the cluster size distribution and
hysteresis.


AP in 2D site percolation is defined as follows: (I) We select
two sites $\alpha$ and $\beta$ at random. (II) Let
$\{s_{\alpha_1},s_{\alpha_2},\cdots,s_{\alpha_n}\}$
($\{s_{\beta_1},s_{\beta_2},\cdots,s_{\beta_m}\})$ be the sizes of
clusters which form into a new big cluster with the size
$\sum_{k=1}^n s_{\alpha_k}+1$ ($\sum_{k=1}^m s_{\beta_k}+1$) by
occupying the site $\alpha$ ($\beta$). Here the cluster size is
defined by the number of sites in the cluster. Then calculate the
products \bea \label{pr} \pi_{\alpha}=\prod_{i=1}^{n}
s_{\alpha_i}~~~\mbox{and}~~~\pi_\beta=\prod_{j=1}^{m} s_{\beta_j}.
\eea This rule is generally called product rule (PR). (III) If
$\pi_\alpha\le \pi_\beta$ ($\pi_\alpha\ge \pi_\beta$) then site
$\beta$ ($\alpha$) remains to be vacant. The processes (II) and
(III) prefer the connection between small clusters, which causes
the cluster repulsion or suppress the growth of large cluster. If
the product in \EQ{pr} is replaced by summation, then the rule is
called as a sum rule. Recent study for site percolation with sum
rule shows that the transition is continuous when the linear size
of the lattice, $L$, goes to infinity \cite{Bastas11}. Since we
use a 2D square lattice $n(m)$ in Eq. (\ref{pr}) is at most 4.

%
\begin{figure}[ht]
\includegraphics[width=8cm]{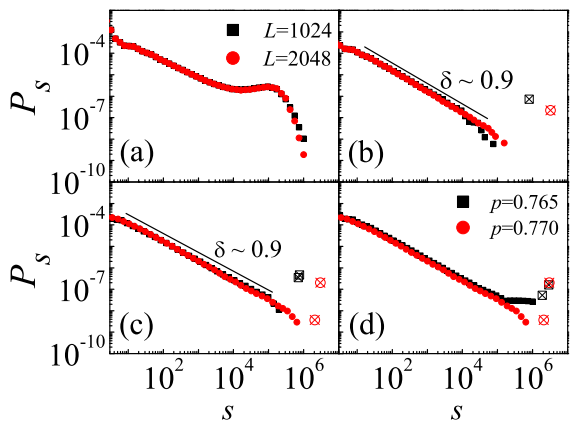}
\caption{(color online) (a) Plot of $P_s$ for $p=0.75$ or $p <
p_c$. The data for $L=1024$ are denoted by black squares and those
for $L=2048$ are denoted by red circles. (b) Plot of $P_s$ for
$p=0.8$ or $p > p_c$. The solid line represents the relation $P_s
\sim s^{-\delta}$ with $\delta\simeq 0.90(2)$. The data points marked
by ``$\times$'' represents the $P_s$ for macroscopically large clusters or
the largest cluster. (c) The same plot for $p=0.77$ or near the
transition threshold. (d) Plot of $P_s$ for $p=0.765$ and $p=0.770$
on the lattice with $L=2048$.} \label{Ps}
\end{figure}
To understand the percolation transition physically to a deeper
level, the properties of the cluster size distribution should be
the first one to understand \cite{Stauffer_book,Lee11,Cho10}.
The cluster size distribution $ \label{defPs} P_s(p)$ at a site
occupation probability $p$ is normally defined by the probability
that a randomly selected site belongs to a cluster which has $s$
sites ($s$-cluster) . For ordinary percolation, it is well known
that $P_s(p_c)$ satisfies a power-law relation
\cite{Stauffer_book},
\bea
\label{ps}
P_s(p_c)\sim s^{-\delta},
\eea
with $\delta\simeq 1.05$ at the percolation transition
probability $p_c$. Since, as we shall show, the percolation
properties under AP depend on the history how the clusters
are grown, we measure $P_s(p)$ by filling sites from the vacant
lattice or increasing $p$.
In \FIG{Ps} $P_s(p)$'s for {\it 2D site percolation under
AP with product rule } ({\bf 2DSAP}) are displayed. $P_s$ in
\FIG{Ps} shows an anomalously unique behavior compared to that of
the {\it ordinary percolation} ({\bf OP}) \cite{Stauffer_book} and
that of the {\it AP percolation on the complete graph} ({\bf
APCG}), which was argued to undergo the continuous phase
transition \cite{Lee11}.

When $p<p_c$, $P_s$ for 2DSAP has a hump in the tail as $p$
approaches to $p_c$ \cite{Cho10}. In this regime, $P_s$ does not
depend on $L$ or $N(=L\times L)$ as shown in \FIG{Ps}(a). In OP,
$P_s$ normally decays exponentially as $s$ gets larger in this
regime. For the detailed comparison to those of OP and APCG, let's
call $s$ at which the hump is maximal $s_H$. In OP we cannot
identify $s_H$. In APCG $s_H$ and $P(s_H)$ was argued to satisfy
the scaling behavior, $s_H \simeq N^{x}$ and $P(S_H) \simeq
N^{-y}$ with $x>0$ and $y>0$ \cite{Lee11}. Therefore in APCG the
hump has the negligible contribution and $P_s$ satisfies the same
scaling form as Eq.(\ref{ps})  in the limit $N \rightarrow
\infty $. This $P_s$ behavior in APCG is believed to be one of the
signals for the continuous transition as in OP. In contrast $s_H$
and $P(s_H)$ of 2DSAP do not depend on $L$ or $s_{H} \simeq const$
and $P(s_H) \simeq const$ as $L$$(N)$ gets larger. We have
numerically checked this behavior for $N=2^{16}, 2^{18}, 2^{20},
2^{22}$. This behavior for 2DSAP means that there should be many
considerably large stable microscopic $s$-clusters with $s\simeq
s_H$ before transition, which indicates the unstable or sudden
appearance of the macroscopic cluster by connecting these clusters
when $p$ increases.

Even when $p> p_c$, $P_s$ for 2DSAP has a unique behavior as shown
in  \FIG{Ps}(b). Except $P_s$ for the macroscopically large
cluster, $P_s$ for microscopically finite clusters for $p > p_c$
still satisfies the same power-law $ P_s = A s^{-\delta} $ with
the same exponent $\delta$ as $P_s(p_c)$ or $\delta =0.90(3)$,
which we will explain with the data in \FIG{Ps}(c). The difference
between $P_s(p>p_c)$ and $P_s(p_c)$ is in $A$ and the tail part
for finite clusters. As $p$ becomes larger than $p_c$, $A$
decreases and the length of tail becomes shorter compared to $P_s$
at $p_c$. The power-law behavior is very robust, because it
maintains for nearly four decades or $10^1 -10^5$ as shown in
\FIG{Ps}(b) before appearing finite-size effects. Moreover, the
power-law behavior for $p>p_c$ is nearly independent of $L$ as for
$p<p_c$ (see \FIG{Ps}(b)).
 This power-law behavior for the finite clusters has been
confirmed even for large $p$ upto $p = 0.9$. In contrast $P_s$ of
microscopic clusters for $p> p_c$ in OP and APCG  exponentially
decays. In 2DSAP the product rule makes the macroscopic
cluster absorb relatively smaller clusters when $p$ gets large in
the regime $p>p_c$. Therefore the larger microscopic clusters
cannot easily disappear. The sustainablity of such metastable
clusters seems to be the origin of the power-law of $P_s$ for
$p>p_c$. As we shall see the hysteresis of 2DSAP is consistent
with the power-law for $p>p_c$ because of such metastable states.

The phase transition for 2DSAP naturally occurs at $p$ which
divides the two regimes of $P_s$ described  by \FIG{Ps}(a) and
\FIG{Ps}(b). The transition threshold $p_c$ for 2DSAP is estimated
by the data in \FIG{Ps}(c) and \FIG{Ps}(d). As shown in
\FIG{Ps}(c) and in \FIG{Ps}(d) at $p=0.770$ $P_s$ for the
macroscopically large cluster starts to be detached from the
continuous distribution of $P_s$ for microscopic clusters. This
detachment behavior seems to be independent of $L(N)$ as
shown in \FIG{Ps}(c). As shown in \FIG{Ps}(d), this detachment
behavior barley occurs and the hump-like tail still exists for
$p=0.765$. We have scrutinized $P_s$ between $0.765 < p < 0.770$,
but the sharp discrimination between the hump-like behavior and
the detachment cannot be made. Such complex behavior mixing the
hump and $P_s$ for the macroscopically large clusters for $p
\simeq p_c$ seems to be a unique behavior of 2DSAP. Therefore the
best estimation of $p_c$ from the numerical data of $P_s$ is
$p_c=0.768(3)$. At $p \simeq p_c$, $P_s$ satisfies the power-law
$P_s = A s^{-\delta}$ with $\delta = 0.90(2)$ very well except for
the very tiny tail part. Again this power-law $P_s =A s^{-0.9}$ is
very robust and holds for more than four decades.
 The result $\delta=0.9$ also provides a very important clue to
understand the transition nature of 2DSAP. Since $P_s$ is a
probability, $P_s$ should satisfy the normalization condition,
$\sum_s P_s=p$. However, the summation $\sum_s^\infty P_s$
diverges if $\delta<1$. Therefore, there should be a cutoff $s_c$
in the upper limit as $\sum_s^{s_c} P(s)=p$. In the limit
$N\rightarrow\infty$, $s_c/N\rightarrow 0$.  Thus, there should be
a discontinuous jump to produce a macroscopic cluster in the limit
$N\rightarrow\infty$ and the transition becomes discontinuous. The
physical origin of the discontinuous transition should come from
the merge of $s$-clusters with $s\simeq s_h$ to form the
macroscopic cluster when $p$ gets larger to be $p=p_c$.


%
\begin{figure}[ht]
\includegraphics[width=7cm]{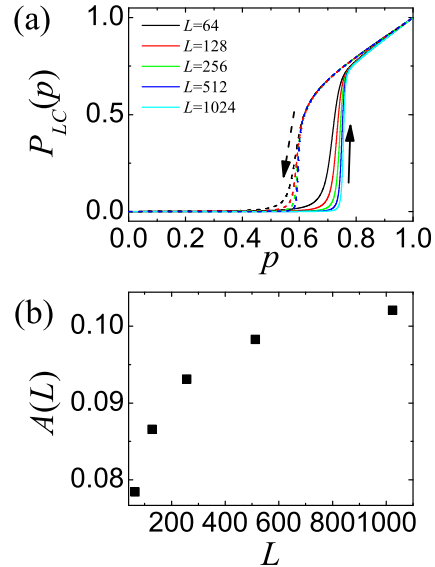}
\caption{(color online) (a) Plot of $P_{LC}(p)$ for the process of
increasing $p$ (solid line) and that for the process of decreasing
$p$ (dashed line). $L$ varies from 64 (left most line) to 1024
(right most line). (b) Plot of the area, $A(L)$, enclosed by
$P_{LC}$.} \label{hysteresis}
\end{figure}

One of the most generally accepted and the simplest methods to
verify whether the observed transition is discontinuous or not is
the measurement of the hysteresis \cite{Landau_book}. The
hysteresis measurement for the explosive percolation has also been
emphasized in Refs.\cite{YKim10,Bastas11}. The hysteresis is a
history-dependent property of a system and usually observed in the
discontinuous phase transition because of the metastable state. If
the transition is discontinuous, then the route of changes in the
order parameter $P_{LC}$ during the process of filling sites from
the vacant lattice or increasing $p$ would be different from that
for the process of deleting sites from the fully-occupied lattice
or decreasing $p$. The order parameter, $P_{LC}$, is defined by
the probability that a site belongs to the largest cluster
\cite{Stauffer_book,Ziff09}; \bea \label{def_P}
P_{LC}=\frac{N_{LC}} {N}. \eea Here, $N_{LC}$ is the number of
sites in the largest cluster. In \FIG{hysteresis}(a), to check the
existence of hysteresis we compare the measured $P_{LC}$'s along
the process of increasing $p$ (solid lines) and along the process
of decreasing $p$ from $N=2^{12}$ to $N=2^{20}$. For the
decreasing process, we slightly modify the rule (III) to easily
break the larger clusters into smaller ones \cite{YKim10},
since the rules (II)
and (III) suppress the formation of a large cluster; i.e., if
$\pi_\alpha\ge \pi_\beta$ then we delete the site $\alpha$. With
this modified rule we find that there exists a hysteresis for
various $L$ as shown in \FIG{hysteresis}(a).

Now the remaining question is whether the hysteresis robustly
remains in the $L\rightarrow\infty$ limit. For the systematic
analysis, we show the dependence of area, $A(L)$, enclosed by
$P_{LC}(L)$ for the increasing and decreasing processes on $L$. If
the system undergoes a continuous transition, then $A(L)$ should
vanish in the limit $L\rightarrow\infty$. However, our data
clearly shows that $A(L)$ increases as $L$ increases or, at least,
seems to saturate to a nonzero value unlike the sum rule
\cite{Bastas11} in which $A(L)\rightarrow 0$ as
$L\rightarrow\infty$. This shows that 2DSAP undergoes a
discontinuous transition. And in 2D lattice, the product rule
makes a completely different transition nature from that of the
sum rule \cite{Bastas11}. This hysteretic property of 2DSAP should
be from the sustainability of the metastablly larger clusters,
which is consistent with the analysis of $P_s$ in \FIG{Ps}.

\begin{figure}[ht]
\includegraphics[width=7cm]{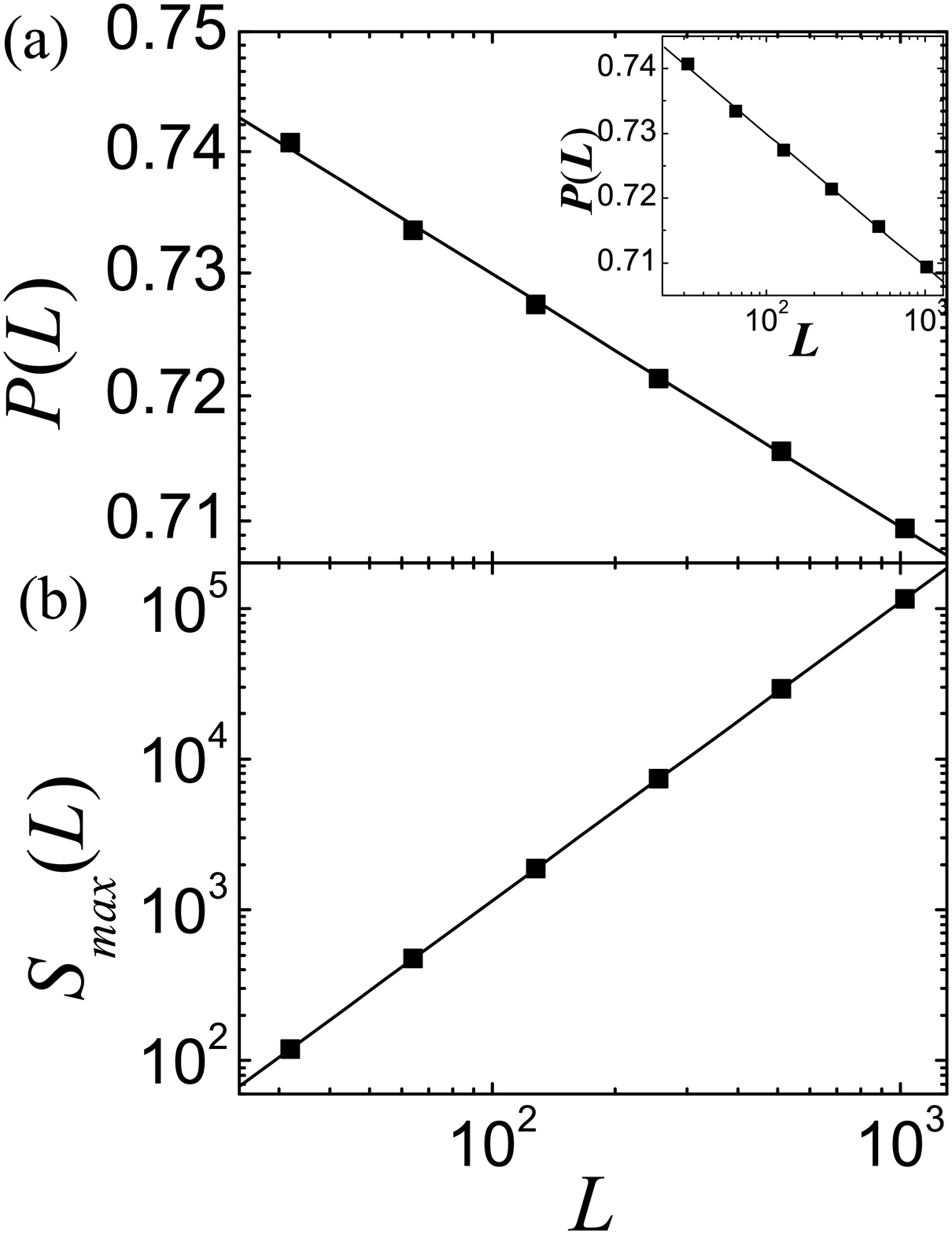}
\caption{(a) Plot of $P_{LC}(L)$ measured at $p_c$ against $L$.
The solid line represents the relation $P(L)\sim L^{-B}$ with
$B=0.012$. Inset: Plot of $P_{LC}(L)$ in semi-log scale. The solid
line represents the relation $P(L)\sim -\log L$. (b) Plot of
$S_{max}(L)$ against $L$. The solid line denotes $S_{max}\sim L^C$
with $C=1.98$.} \label{betagamma}
\end{figure}

Since we don't know the physically corresponding formula to
Hamiltonian or free energy for 2DSAP and there exists the
nontrivial hysteretic property, it might be physically nonsense to
discuss about the finite-size scaling. However for the
comparison's purpose to other works on explosive percolation
\cite{Ziff09,Cho09,Radicchi}, we now present the finite size
analysis around $p_c=0.768(3)$, which is the percolation
transition probability for the $p$-increasing process. From the
data in \FIG{hysteresis}(a), $P_{LC} (L)$ at the $p_c$ is
estimated as shown in \FIG{betagamma}(a). $P_{LC} (L)$ seems to
satisfy the relation $P(L)\sim L^{-B}$ with $B=0.011(2)$, where
conventionally $B$ corresponds to $\beta/\nu$. This value of $B$
is very close to zero. Thus, in the inset of \FIG{betagamma}(a),
we also fit the data to the relation $P(L)\sim -\log L$ which
corresponds to the case $B \rightarrow 0$. Since we cannot exclude
the possibility $B=0$ or $\beta=0$, the possibility for $P_{LC} (L
\rightarrow \infty)$ at $p_c$ to have discontinuous jump cannot be
excluded. We also measure the mean cluster size, defined by \bea
\label{def_S} S(p,L)=\frac{\sum_s^{\prime} sP_s}{\sum_s^{\prime}
P_s}. \eea $\sum_s^{\prime}$ represents the summation over all $s$
except the largest one.  $S(p,L)$'s maximal value, $S_{max}(L)$,
is displayed in \FIG{betagamma}(b). Again we fit the data to the
conventional scaling relation $S_{max}(L)\sim L^{-C}$, and we
obtain $C=1.98(1)$, where $C$ corresponds to conventional
$\gamma/\nu$.
%


In summary, we study the site percolation under AP with a product
rule in a 2D lattice. From the measurement of $P_s(p)$, we find
that the $P_s(p)$ have a very stable hump when $p<p_c$. This
indicates that below $p_c$ large number of stable $s$-clusters
with $s\simeq s_H$ exist but their sizes are still microscopic.
As $p$ approaches to $p_c$, $P_s(p)$ has a very robust power-law
regime followed by the hump.
Since the obtained value of the exponent, $\delta$, for the
power-law regime in the vicinity of $p_c$ is less than unity,
there should be a cutoff $s_c$ in the possible cluster size for
$p\simeq p_c$ unlike OP \cite{Stauffer_book}. Thus, to generate a
macroscopic cluster there should be a discontinuous jump in the
limit $L\rightarrow\infty$ and the transition becomes
discontinuous. The non-vanishing  hysteresis in $P_{LC}$
also verifies that the transition is
discontinuous. This result clearly shows that the percolation
transition caused by the product rule in a 2D square lattice is
discontinuous.

This work was supported by National Research Foundation of Korea
(NRF) Grant funded by the Korean Government (MEST) (No.
2009-0073939 and No. 2011-0015257).

\end{document}